\documentclass[3p,times,procedia]{elsarticle}
\usepackage{nupha_ecrc}

\volume{00}

\firstpage{1}

\journalname{Nuclear Physics A}

\runauth{}

\jid{nupha}

\jnltitlelogo{Nuclear Physics A}

\usepackage{amssymb}

\usepackage[figuresright]{rotating}

\RequirePackage{doi}
\usepackage{hyperref}

\begin{document}

\begin{frontmatter}



\dochead{XXVIIIth International Conference on Ultrarelativistic Nucleus-Nucleus Collisions\\ (Quark Matter 2019)}

\title{Jet splitting measurements in Pb--Pb and pp collisions at $\sqrt{s}_{\mathrm{NN}} =$ 5.02 TeV with ALICE}


\author{Laura Havener 
 on behalf of the ALICE collaboration}

\address{Yale University}

\begin{abstract}

Recent ALICE measurements of jet splittings in Pb--Pb and pp collisions at $\sqrt{s_{\mathrm{NN}}}$ = 5.02 TeV are reported. These measurements scan the phase space of jet emissions in search of medium-induced signals which are expected to emerge at different scales. These include effects such as multiple soft-radiation, single hard emissions, and color coherence. The Lund plane diagram is shown, including projections onto distributions of the splitting scale $k_{\mathrm{T}}$ in intervals of the splitting angle $R_{\mathrm{g}}$. Soft Drop grooming is applied to access hard splittings within the jet, enabling measurements of groomed substructure variables. These include the shared momentum fraction $z_{\mathrm{g}}$ between the two hardest subjets and the number of Soft Drop splittings $n_{\mathrm{SD}}$. 
The results in Pb--Pb collisions are compared to PYTHIA events embedded into a Pb--Pb background to separate out background from in-medium effects. Measurements of $z_{\mathrm{g}}$ and the normalized splitting angle $\theta_{\mathrm{g}}$ will also be shown in pp collisions at $\sqrt{s}$ = 5.02 TeV for different grooming settings.

\end{abstract}

\begin{keyword}

Nuclear Physics \sep Heavy-Ion Collisions  \sep Quark-Gluon Plasma \sep QCD \sep Jets \sep Jet Quenching \sep Jet Substructure 


\end{keyword}

\end{frontmatter}


\section{Introduction}
\label{sec:intro}
Jet substructure measurements investigate how the internal structure of a jet is modified by the medium produced in heavy-ion collisions. Specifically, they probe the phase space of jet emissions in search of medium-induced signals from effects such as color coherence, multiple soft radiation, and single hard emissions using the Lund plane~\cite{Dreyer:2018nbf, Andrews:2018jcm}. The Lund plane is a 2D diagram in $\ln{k_{\mathrm{T}}}$ and $\ln{1/\Delta R}$, where $k_{\mathrm{T}}$ is the relative transverse momentum and $\Delta{R} = \sqrt{(\eta_{1} - \eta_{2})^{2} + (\phi_{1} - \phi_{2})^{2}}$ is the angular distance between two subjets 1 and 2. The Lund diagram is constructed experimentally by finding jets using the anti-$k_{t}$ algorithm~\cite{Cacciari:2008gp}, re-clustering them with the Cambridge/Aachen (C/A) algorithm~\cite{Dokshitzer:1997in} to enforce angular ordering, and filling the diagram with information about the first hard splittings. It can be used to separate out different types of jets, including jets less impacted by the background, through grooming procedures. 

Soft Drop (SD) grooming~\cite{Larkoski:2014wba} is used to access hard splittings within a jet to measure groomed substructure variables. This procedure finds the first splitting in a declustered C/A jet that passes a grooming condition on the shared transverse momentum ($p_{\mathrm{T}}$) fraction between the subjets $z = \frac{\mathrm{min}(p_{\mathrm{T1}},\: p_{\mathrm{T2}})}{p_{\mathrm{T1}}+p_{\mathrm{T2}}}$, where $p_{\mathrm{T1}}$ and $p_{\mathrm{T2}}$ are the higher and lower $p_{\mathrm{T}}$ subjets, respectively. The SD grooming condition is defined as $z > z_{\mathrm{cut}}\left(\frac{R_{\mathrm{g}}}{R_{0}}\right)^{\beta}$, where $R_{\mathrm{g}}$ is the $\Delta{R}$ for the subjets passing SD, $R_{0}$ is the jet radius, and $\beta$ and $z_{\mathrm{cut}}$ are tuneable parameters with default values of 0 and 0.1, respectively.
The $z$ and $k_{\mathrm{T}}$ of the groomed jets are denoted $z_{\mathrm{g}}$ and $k_{\mathrm{Tg}}$, where smaller and larger $z_{\mathrm{g}}$ values correspond to more asymmetric and symmetric splittings, respectively. The $R_{\mathrm{g}}$ (or $\theta_{\mathrm{g}} = R_{\mathrm{g}}/R_{0}$) separates wider and narrower angle splittings which have different formation times with wide splittings forming earlier and narrow splittings forming later. The earlier emissions could see more of the medium and experience more modification. This measurement also uses iterative SD which follows the hardest branch to determine which splittings pass SD instead of stopping at the first, where $n_{\mathrm{SD}}$ is the number of splittings that pass SD as the splittings are unwound. 
ALICE previously measured jet substructure in Pb--Pb collisions at $\sqrt{s_{\mathrm{NN}}} =$ 2.76 TeV~\cite{Acharya:2019djg}. The new results presented here at $\sqrt{s_{\mathrm{NN}}} =$ 5.02 TeV benefit from a substantial increase in the dataset size, allowing for more differential studies. 

\section{Analysis}
\label{sec:analysis}

This analysis uses 2018 0--10\% Pb--Pb and 2017 pp collision data from the ALICE detector~\cite{Aamodt:2008zz} at $\sqrt{s_{\mathrm{NN}}} = 5.02$ TeV. The Pb--Pb data is not unfolded for detector effects, but is compared to PYTHIA8~\cite{Sjostrand:2007gs} (Monash Tune) Monte Carlo (MC) simulations embedded into real Pb--Pb data such that the MC has the same background as the data. 
The pp data is unfolded for detector effects using 2D Bayesian unfolding~\cite{RooUF} with a response built from PYTHIA8 generated jets that were propagated through a GEANT3 simulation~\cite{Brun:1119728} of the ALICE detector. Track-based jets are reconstructed from tracks with $p_{\mathrm{T}} > 150$ MeV/$c$ using the anti-$k_{\mathrm{T}}$ algorithm with $R=0.4$. Jets in Pb--Pb collisions have a large background contribution that is removed through the jet-by-jet constituent subtraction method~\cite{Berta:2014eza}.  

\section{Results}
\label{sec:results}

\begin{figure}[h!]
\begin{center}
    \includegraphics[width=0.46\textwidth]{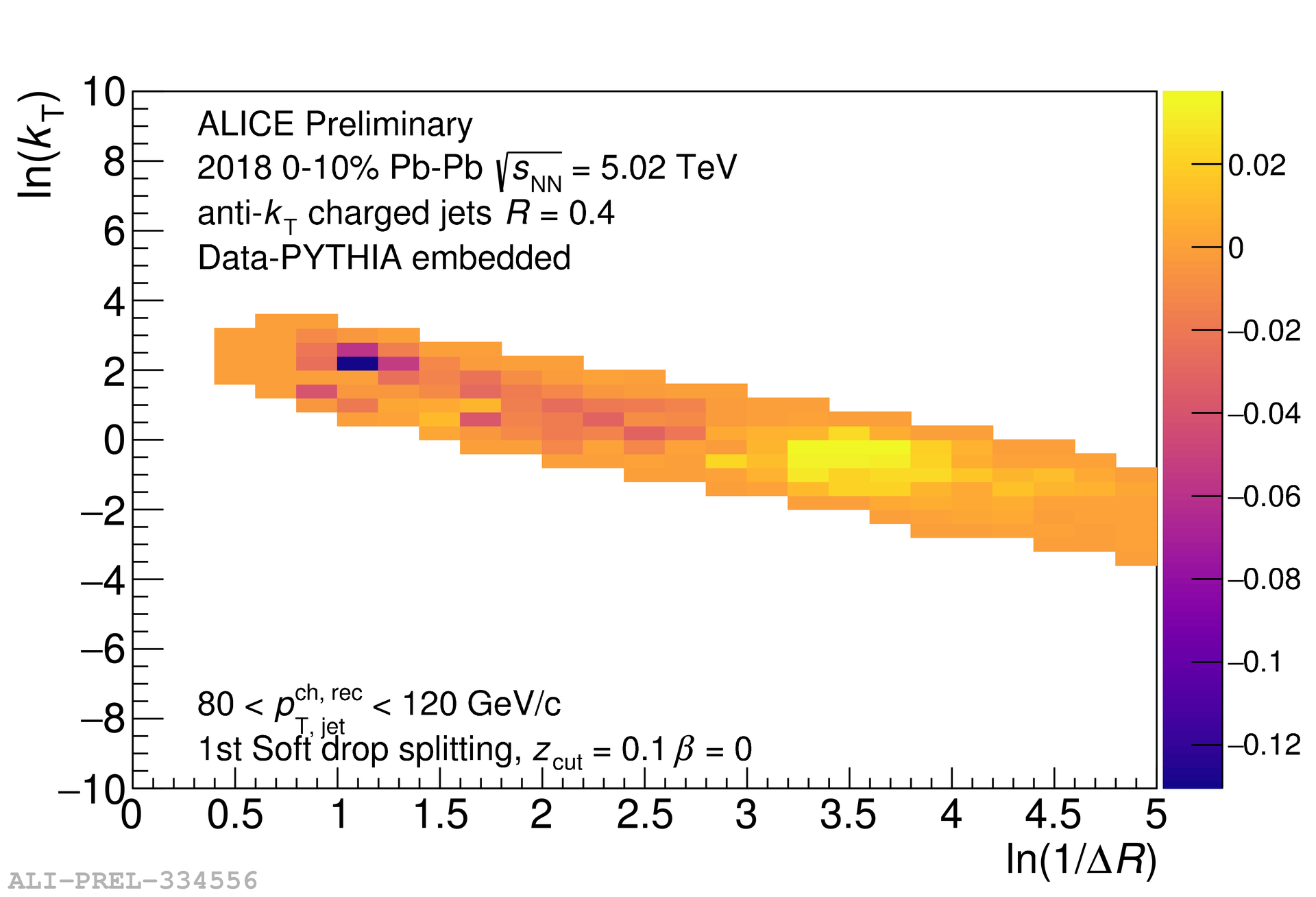}
    \includegraphics[width=0.46\textwidth]{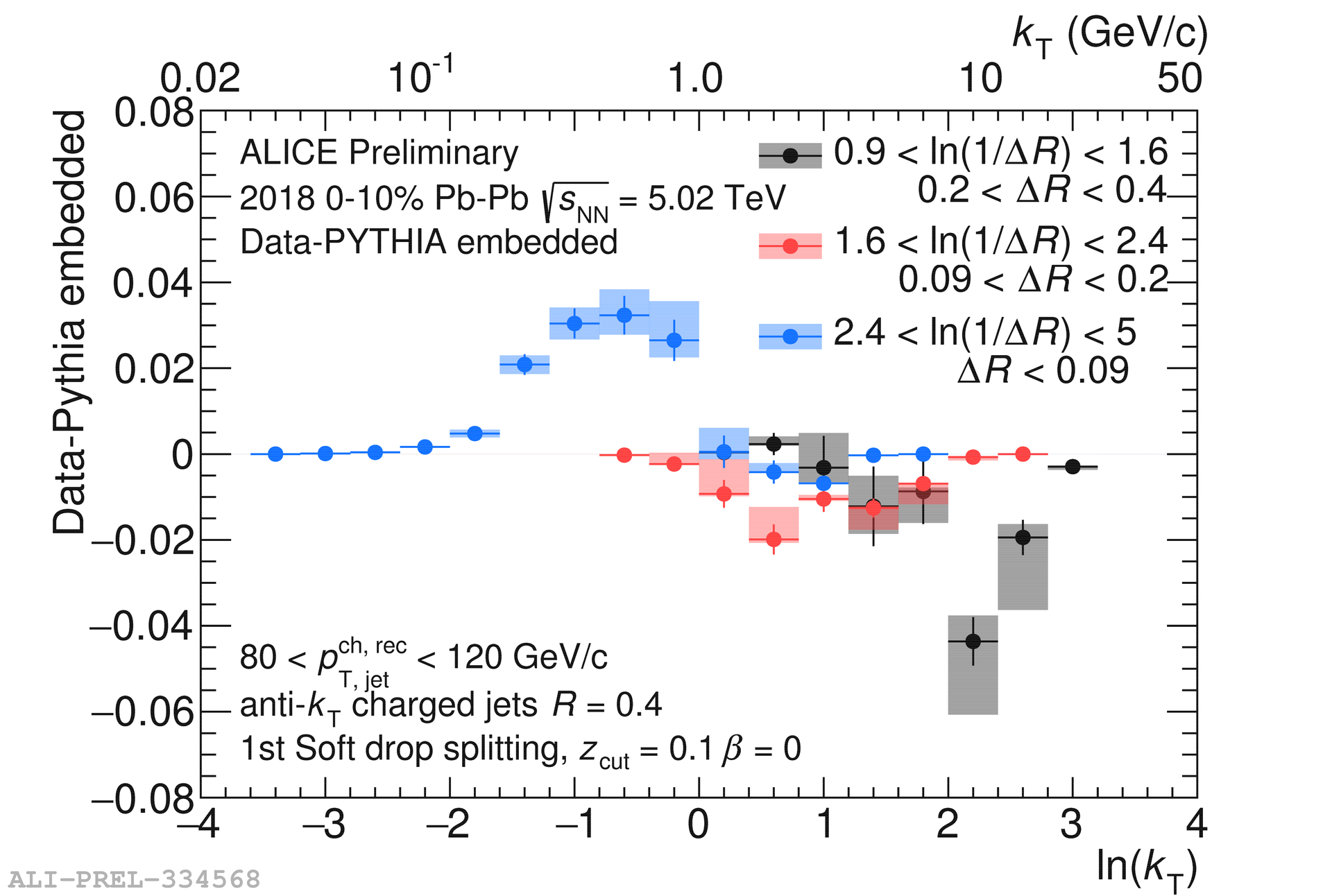}
    \caption{Left: The difference of the Lund plane ($\ln{k_{\mathrm{T}}}$ vs. $\ln{1/\Delta R}$) distributions for 0--10\% Pb--Pb data and embedded MC for track-based jets in the $p_{\mathrm{T}}$ range 80--120 GeV/$c$. The distributions are normalized to the number of jets in the $p_{\mathrm{T}}$ range 80--120 GeV/$c$ before the subtraction is performed such that the z-axis is the difference in the per jet yield. Right: Projections onto the $\ln{k_{\mathrm{T}}}$ axis in intervals of $\ln{1/\Delta R}$.}
    \label{fig:lunddataembed}
\end{center}
\end{figure}

The Lund Plane is measured in 0--10\% Pb--Pb collisions and is compared to embedded MC with SD grooming applied. The distributions are subtracted from each other to remove additional background effects and is shown on the left in Fig.~\ref{fig:lunddataembed} for jets in the $p_{\mathrm{T}}$ range 80--120 GeV/$c$. 
Both distributions are normalized to the total number of jets in the $p_{\mathrm{T}}$ range 80--120 GeV/$c$ before subtraction, such that anything above or below zero represents an enhancement or a suppression. A suppression is seen at large $\Delta R$ (or small $\ln{1/\Delta R}$) with a corresponding enhancement at small $\Delta R$. The right panel in Fig.~\ref{fig:lunddataembed} shows the projections onto the $\ln{k_{\mathrm{T}}}$ axis in intervals of $\ln{1/\Delta R}$ (motivated in Ref~\cite{Dreyer:2018nbf}). The smaller $\Delta R$ intervals show an enhancement in the non-perturbative regime ($\ln{k_{\mathrm{T}}} < 0$) and the larger $\Delta R$ intervals show a suppression. 

\begin{figure}[h!]
\begin{center}
    \includegraphics[width=0.37\textwidth]{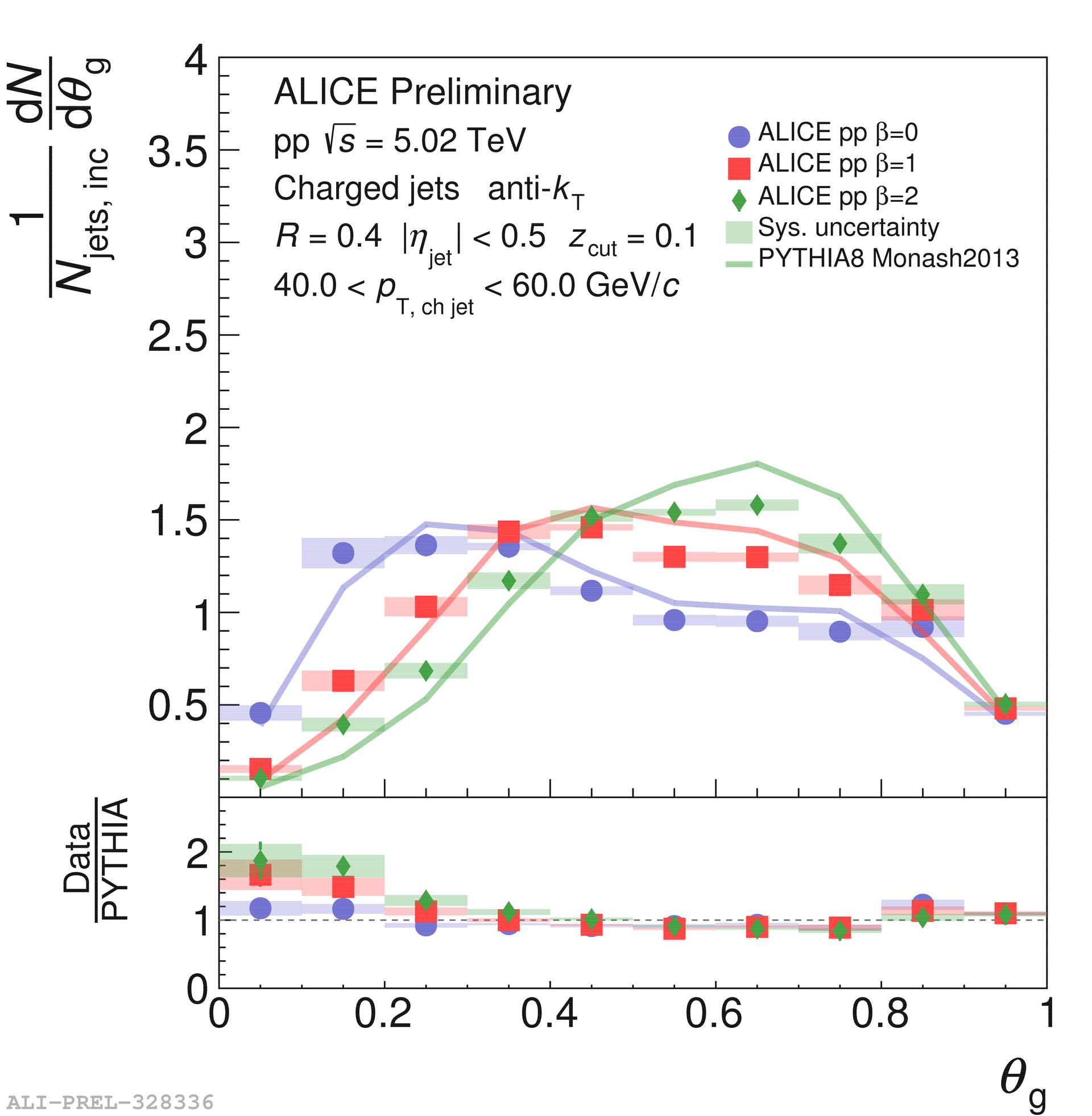}
    \includegraphics[width=0.37\textwidth]{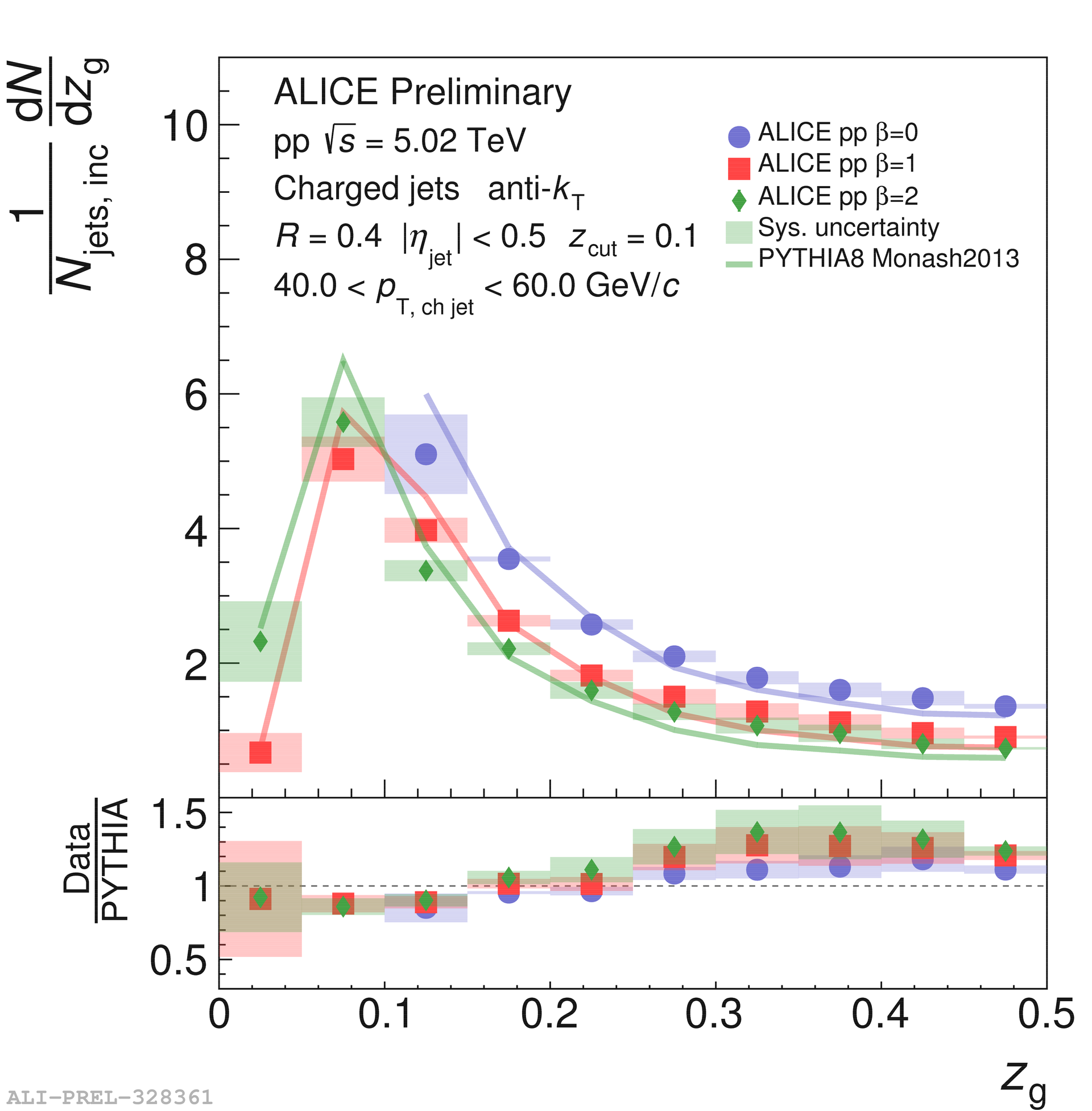}
     \caption{The fully corrected $\theta_{\mathrm{g}}$ (left) and $z_{\mathrm{g}}$ (right) distribution for track-based jets in the $p_{\mathrm{T}}$ range 
     40--60 GeV/$c$ 
     in pp collisions at $\sqrt{s}$ = 5.02 TeV for different grooming conditions. Both the data (markers) and PYTHIA8 simulations (lines) are shown with the ratio of data to MC in the bottom panel. The distributions are normalized to the number of jets in the $p_{\mathrm{T}}$ range 40--60 GeV/$c$. }
\label{fig:pp}
\end{center}
 \end{figure}

The fully corrected $\theta_{\mathrm{g}}$ and $z_{\mathrm{g}}$ distributions in pp collisions are shown in Fig.~\ref{fig:pp} for jets in the $p_{\mathrm{T}}$ range 40--60 GeV/$c$. 
This is shown for different grooming settings by varying $\beta$ which is useful for constraining pQCD calculations and non-perturbative effects~\cite{Kang:2019prh}. Increasing $\beta$ increases the contribution of jets with wider angle, more asymmetric splittings. All the distributions are shown to be mostly consistent with PYTHIA8 Monash MC simulations through the ratio in the bottom panel. These fully corrected pp measurements will serve as a baseline for future unfolded Pb--Pb measurements. 

\begin{figure}[h!]
\begin{center}
    \includegraphics[width=0.49\textwidth]{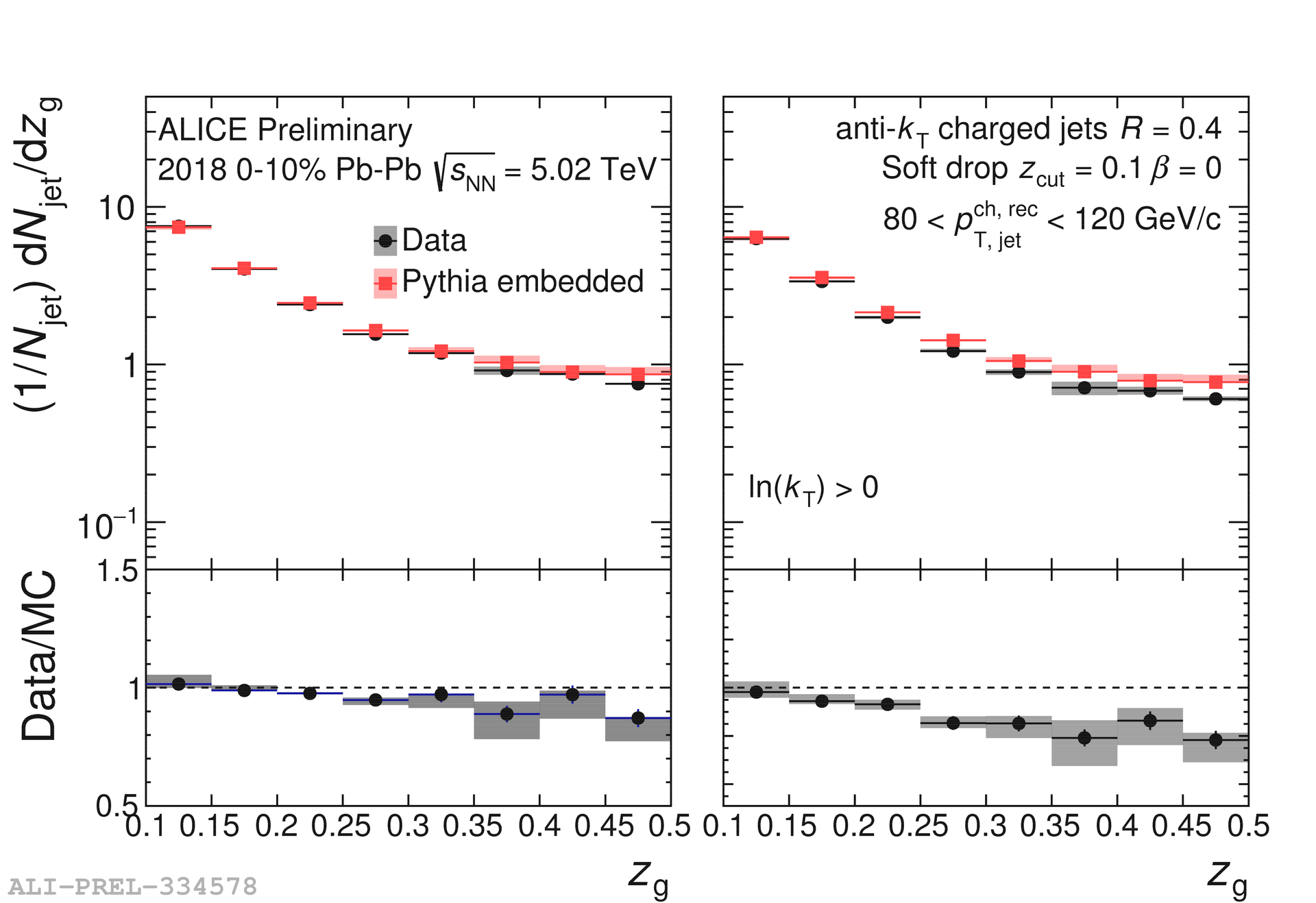}
    \includegraphics[width=0.49\textwidth]{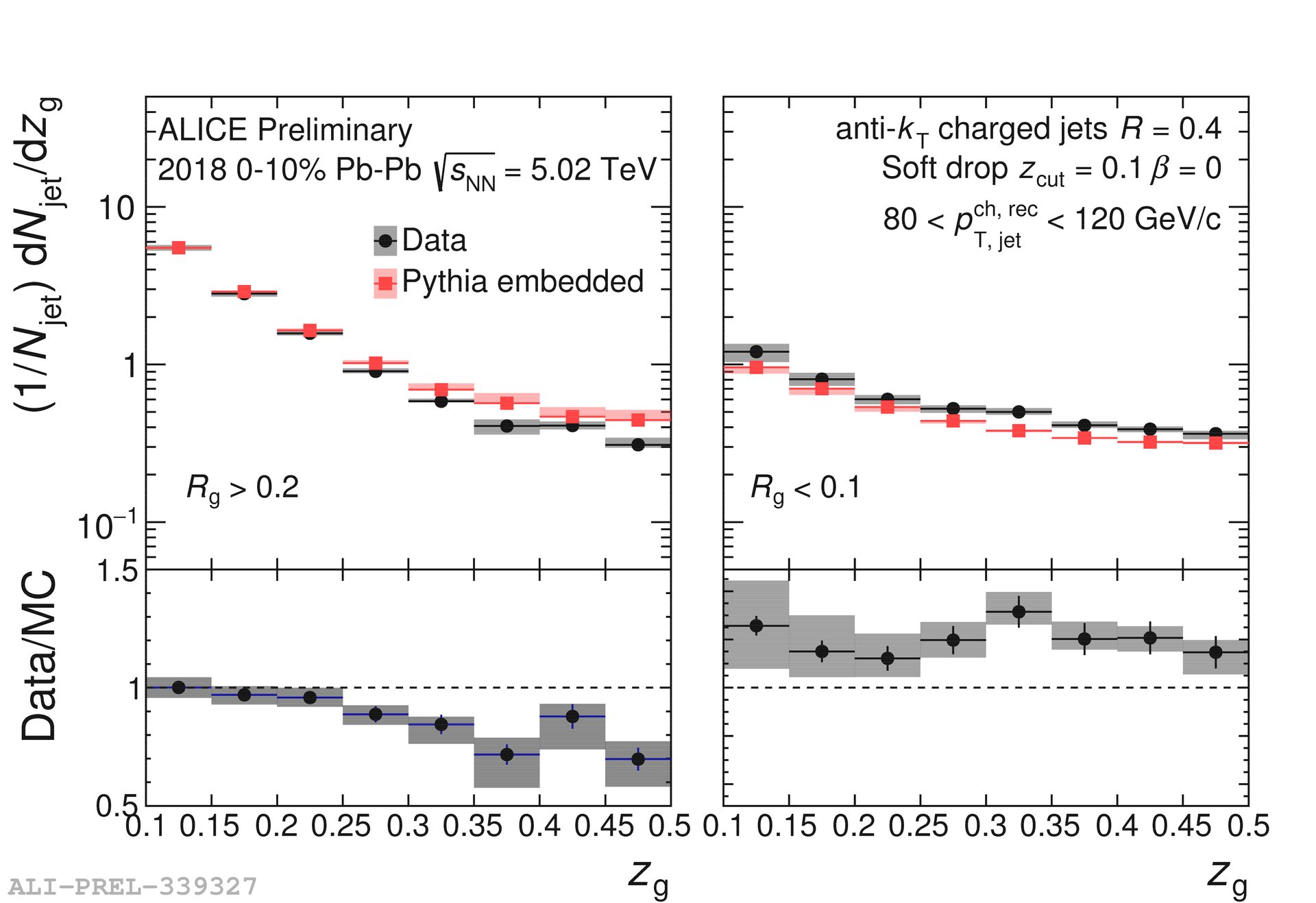}
    \caption{The $z_{\mathrm{g}}$ distribution for track-based jets in the $p_{\mathrm{T}}$ range 80--120 GeV/$c$ in 0--10\% Pb--Pb data in black compared to embedded MC in red. The distributions are normalized to the total number of jets in the $p_{\mathrm{T}}$ range 80--120 GeV/$c$, not just the ones that pass SD.  The far left panel is for all splittings, the center left panel is for splittings with $\ln{k_{\mathrm{T}}} > 0$, the center right is for wider splittings with $R_{\mathrm{g}} > 0.2$, and the far right is for narrower splittings with $R_{\mathrm{g}} < 0.1$. The bottom panels are the ratios of the data to the embedded MC.  }
    \label{fig:zg}
\end{center}
\end{figure}

Unfolding substructure variables is challenging in Pb--Pb collisions due to the large uncorrelated background that leads to a large contribution from incorrect splittings. Therefore, embedded MC is used as a reference such that background effects are effectively canceled in any comparisons. The left-most panel of Fig.~\ref{fig:zg} shows the $z_{\mathrm{g}}$ distribution in 0--10\% Pb--Pb collisions compared to embedded MC, with the ratio in the bottom panel.
The ratio shows a slight suppression of more symmetric splittings. A selection on the $k_{\mathrm{Tg}}$ of the splittings greater than 1 GeV/$c$ (or $\ln{k_{\mathrm{Tg}}} > 0$) is applied to remove non-perturbative splittings~\cite{Cunqueiro:2018jbh}. This is shown in the center left panel of Fig.~\ref{fig:zg}, where a more significant suppression is observed. The two right panels show the $z_{\mathrm{g}}$ distributions for different cuts on $R_{\mathrm{g}}$, where wider splittings are shown on the center right ($R_{\mathrm{g}} > 0.2$) and narrower splittings on the far right ($R_{\mathrm{g}} <0.1$).  The wider splittings have a more significant suppression than the inclusive jets and the narrower splittings have a corresponding enhancement. 

\begin{figure}[h!]
\begin{center}
    \includegraphics[width=0.58\textwidth]{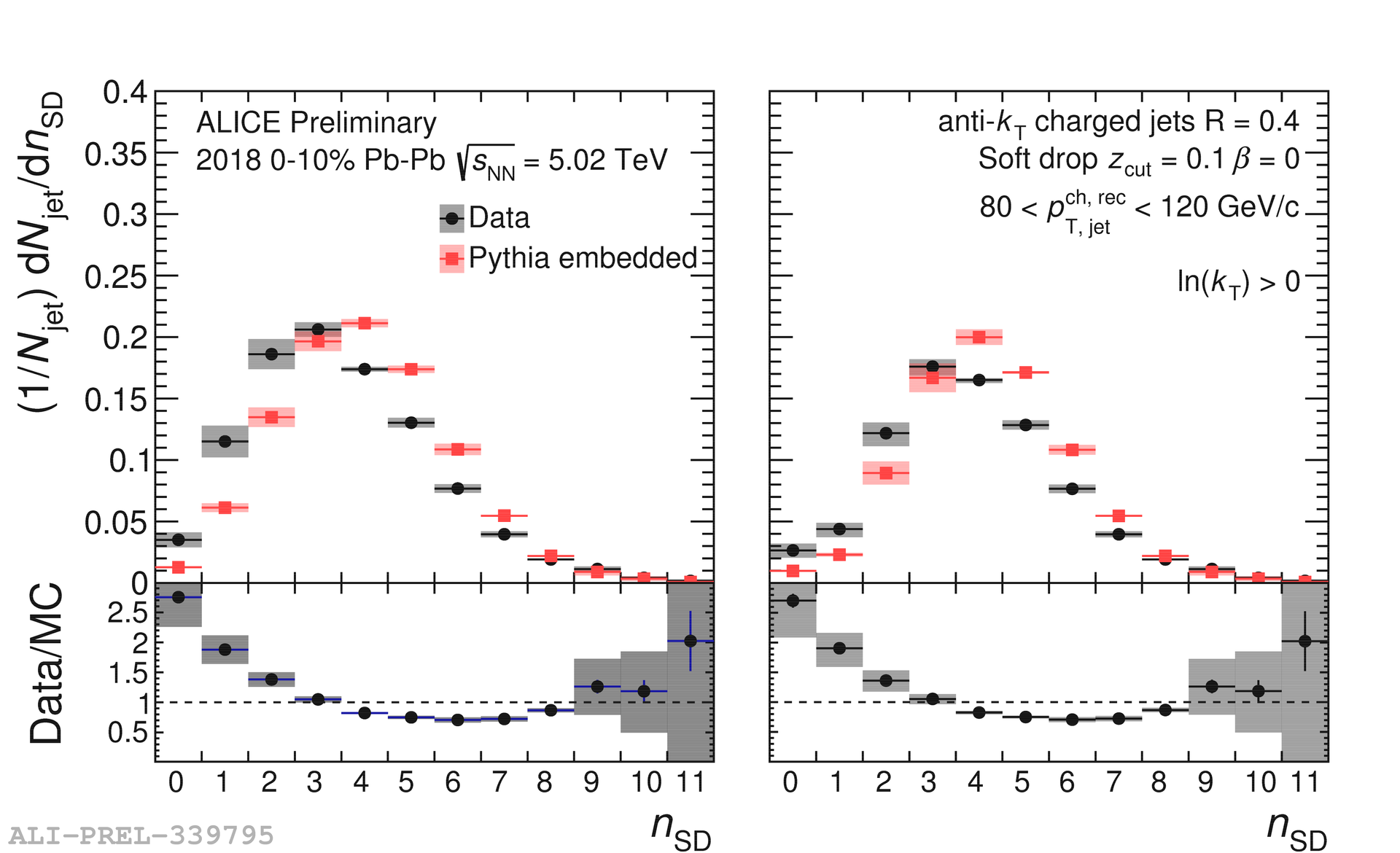}
    \caption{The $n_{\mathrm{SD}}$ distribution for Pb--Pb data in black and enbedded MC in red for 0--10\% centrality and track-based jets in the $p_{\mathrm{T}}$ range 80--120 GeV/$c$. The distributions are normalized to the number of jets in the $p_{\mathrm{T}}$ range 80--120 GeV/$c$. The left panel is for all jets and the right panel is for jets with $\ln{k_{\mathrm{T}}} > 0$. The bottom panel is the ratio of the data to the MC.}
    \label{fig:nSD}
\end{center}
\end{figure}

The $n_{\mathrm{SD}}$ distributions for jets in the $p_{\mathrm{T}}$ range 80--120 GeV/$c$ in Pb--Pb collisions compared to embedded MC are shown in Fig.~\ref{fig:nSD} for all splittings on the left and splittings with $\ln{k_{\mathrm{T}}} > 0$ on the right. The ratio of the data to the embedded MC is shown in the bottom panel. In both cases the distribution is significantly modified with an enhancement at lower $n_{\mathrm{SD}}$ values and a suppression at higher $n_{\mathrm{SD}}$ values. 

\section{Conclusions and Outlook}
\label{sec:conclusion}
New measurements of fully corrected $z_{\mathrm{g}}$ and $\theta_{\mathrm{g}}$ distributions are shown in pp collisions at $\sqrt{s} =$ 5.02 TeV. They are mostly consistent with MC simulations and will serve as a baseline for future unfolded measurements in Pb--Pb collisions. New measurements of $z_{\mathrm{g}}$ in both the perturbative and non-perturbative regime in intervals of the splitting angle for 0--10\% Pb--Pb collisions at $\sqrt{s_{\mathrm{NN}}} =$ 5.02 TeV are also shown. A suppression of the larger angle splittings in more perturbative regions with a corresponding enhancement of narrower angle splittings in more non-perturbative regions is observed. This is consistent with the idea that wider splittings are formed earlier and thus traverse more of the medium. Therefore, wider splittings should be more affected by the medium than narrower splittings and thus experience more modification. The $n_{\mathrm{SD}}$ distributions show a significant modification in Pb--Pb collisions with an enhancement of smaller values and a suppression of larger values. 
The next step is to investigate methods to suppress the background splittings in Pb--Pb collisions such that the substructure variables can be unfolded for detector effects and compared directly to theoretical predictions in order to further constrain jet quenching models. 


\bibliographystyle{elsarticle-num}
\bibliography{mybib}







\end{document}